\documentclass[%
 reprint,
superscriptaddress,
 amsmath,amssymb,
 aps,
pra
]{revtex4-2}

\usepackage{graphicx}% Include figure files
\usepackage{dcolumn}% Align table columns on decimal point
\usepackage{bm}% bold math
\usepackage{amsmath}
\usepackage{extpfeil}
\usepackage{hyperref}% add hypertext capabilities
\usepackage{tabularx}  % 支持自动调整列宽
\usepackage{booktabs}  % 用于创建漂亮的水平线
\usepackage{multirow}   % 用于合并多行
\usepackage[normalem]{ulem}

\hypersetup{
    colorlinks=true,
    linkcolor=blue,
    filecolor=blue,      
    urlcolor=blue,
    citecolor=cyan,
}
\begin{document}

\preprint{APS/123-QED}

\title{Superresolution of unequal-brightness thermal sources for stellar interferometry}% Force line breaks with \\

\author{Chenyu Hu}
\affiliation{National Laboratory of Solid State Microstructures, Key Laboratory of Intelligent Optical Sensing and Manipulation,\\
College of Engineering and Applied Sciences, Jiangsu Physical Science Research Center,\\
and Collaborative Innovation Center of Advanced Microstructures, Nanjing University, Nanjing 210093, China}
\author{Ben Wang}
\affiliation{National Laboratory of Solid State Microstructures, Key Laboratory of Intelligent Optical Sensing and Manipulation,\\
College of Engineering and Applied Sciences, Jiangsu Physical Science Research Center,\\
and Collaborative Innovation Center of Advanced Microstructures, Nanjing University, Nanjing 210093, China}
\author{Jiandong Zhang}
\affiliation{National Laboratory of Solid State Microstructures, Key Laboratory of Intelligent Optical Sensing and Manipulation,\\
College of Engineering and Applied Sciences, Jiangsu Physical Science Research Center,\\
and Collaborative Innovation Center of Advanced Microstructures, Nanjing University, Nanjing 210093, China}
\author{Kunxu Wang}
\affiliation{National Laboratory of Solid State Microstructures, Key Laboratory of Intelligent Optical Sensing and Manipulation,\\
College of Engineering and Applied Sciences, Jiangsu Physical Science Research Center,\\
and Collaborative Innovation Center of Advanced Microstructures, Nanjing University, Nanjing 210093, China}
\author{Huigen Liu}
\email{huigen@nju.edu.cn}
\affiliation{School of Astronomy and Space Science, Nanjing University, Nanjing 210093, China,\\
Key Laboratory of Modern Astronomy and Astrophysics, Ministry of Education, Nanjing 210023, China}
\author{Jilin Zhou}
\email{zhoujl@nju.edu.cn}
\affiliation{School of Astronomy and Space Science, Nanjing University, Nanjing 210093, China,\\
Key Laboratory of Modern Astronomy and Astrophysics, Ministry of Education, Nanjing 210023, China}
\author{Lijian Zhang}
\email{ lijian.zhang@nju.edu.cn}
\affiliation{National Laboratory of Solid State Microstructures, Key Laboratory of Intelligent Optical Sensing and Manipulation,\\
College of Engineering and Applied Sciences, Jiangsu Physical Science Research Center,\\
and Collaborative Innovation Center of Advanced Microstructures, Nanjing University, Nanjing 210093, China}

\date{\today}% It is always \today, today,

\begin{abstract}
Resolving high-contrast targets is a fundamental yet highly challenging task in astronomy.  Using quantum estimation theory, we demonstrate that the ultimate limit for estimating the separation between two unequal-brightness thermal sources via interferometry remains constant, enabling the potential for superresolution. We give a comparative analysis of two primary stellar interferometric schemes: amplitude interferometry and intensity interferometry.  Notably, the nulling strategy employed in amplitude interferometry, a configuration specifically proposed for exoplanet detection by leveraging destructive interference to suppress the brighter source, is quantum optimal for separation estimation.  While intensity interferometry is less effective than amplitude interferometry in lossless scenarios and fails to achieve superresolution, it becomes competitive when optical loss in large-scale interferometry is considered. By applying these methodologies to modern stellar interferometry, we highlight the promise of large-scale interferometry for advancing high-resolution astronomical observation.
\end{abstract}

\maketitle

%\tableofcontents

\section{\label{sec:level1}Introduction}
It was long believed that the optimal resolution in imaging is governed by the Rayleigh limit, and it was posited that two point sources are unresolvable when their images overlap excessively \cite{rayleigh1879xxxi}. Recent investigations by Tsang and coworkers have revisited this subject using the quantum estimation theory, i.e., the quantum Fisher information (QFI) and quantum Cram\'{e}r-Rao bound \cite{liu2020quantum}. They showed a constant QFI for the separation between two identical thermal point sources, revealing that the Rayleigh limit is not an ultimate limit \cite{tsang2016quantum,tsang2019resolving}.  This surprising result catalyzed several quantum-inspired superresolution experimental implementations \cite{tang2016fault,paur2016achieving,santamaria2023spatial,tan2023quantum} and has since been generalized to cover more types of sources and imaging modalities. Within conventional imaging, it was subsequently generalized to encompass incoherent point sources with unequal brightnesses \cite{vrehavcek2017multiparameter,vrehavcek2018optimal,prasad2020quantum,wang2021quantum,zhang2024superresolution,santamaria2024single} and thermal fluctuation \cite{nair2016far,lupo2016ultimate}. These results indicate that the ultimate resolution remains tied to the system's aperture. Given the challenges in manufacturing large optical telescopes, interferometers with distributed telescopes, known as aperture synthesis, offer a compelling alternative for high-resolution astronomy by increasing the effective aperture through extended baselines.

Interferometric imaging aims to extract information from the complex degree of coherence (CDC) between telescopes to reconstruct the intensity distribution of the source \cite{goodman2015statistical}.  Related to this purpose, the majority of modern interferometers employ amplitude interferometry, enabling access to the complex CDC by coherently combining photons arriving at different telescopes via optical links \cite{glindemann2011principles,david2015practical}. Theoretical and experimental evidence has established that amplitude interferometry is optimal for estimating the CDC \cite{pearce2017optimal,howard2019optimal}. Specific to resolving point sources, the ultimate resolution limit in the single-photon regime has been well characterized \cite{lupo2020quantum,zanforlin2022optical,sajjad2024quantum,fiderer2021general}. However, real-world scenarios demand consideration of more general photon statistics, as natural emitters are inherently thermal. While amplitude interferometry has been shown to achieve superresolution for two identical thermal sources \cite{wang2021superresolution}, its applicability to sources with unequal brightnesses remains to be fully explored. On the other hand, the efficacy of amplitude interferometry is constrained by decoherence arising from cumulative optical loss \cite{townsend1993single} and random atmospheric fluctuations \cite{fried1966optical}, which limit its performance and restrict the baseline length to a few hundred meters \cite{glindemann2011principles,david2015practical}. To extend baselines, an alternative scheme known as intensity interferometry was developed by Hanbury Brown and Twiss (HBT) \cite{brown1974intensity}. This scheme measures the intensity correlation between telescopes, where optical loss and atmospheric turbulence are less critical concerns since the signals are connected through electronic links rather than optical links \cite{zampieri2021stellar,matthews2023intensity}. This allows for implementation over significantly greater distances in principle. However, it primarily measures the amplitude of the CDC, necessitating phase-recovery algorithms for imaging \cite{liu2025active}. Another downside of intensity interferometry lies in the postselection of photon coincidence events, which results in a low signal-to-noise ratio under conditions of low mode occupancy of thermal light. While kilometer-scale baselines may compensate for the aforementioned limitations \cite{dravins2013optical,kieda2019astro2020,bojer2022quantitative}, it has hitherto remained unclear whether amplitude or intensity interferometry provides superior resolution.

In this paper, we generalize previous studies on single-aperture imaging \cite{vrehavcek2017multiparameter,vrehavcek2018optimal,prasad2020quantum,wang2021quantum,zhang2024superresolution,santamaria2024single} and equal-brightness sources \cite{lupo2020quantum,zanforlin2022optical,sajjad2024quantum,wang2021superresolution} to interferometric systems with unequal-brightness thermal sources. We derive analytical expressions for the quantum-limited precision of separation estimation, as well as the achievable precision using amplitude and intensity interferometry. The results identify optimal measurement strategies to achieve the ultimate limit, which encompasses the nulling strategy—a configuration of amplitude interferometry specifically designed for resolving high-contrast targets, with notable applications in exoplanet detection \cite{glindemann2011principles}.  With the same baseline lengths and negligible optical loss, intensity interferometry performs worse than amplitude interferometry.  However, it shows the potential to surpass amplitude interferometry when significant optical transmission loss is considered.  Finally, we present a quantitative comparison of state-of-the-art amplitude and intensity interferometers for the benchmark problem of determining the separation of two adjacent stars in astronomical observations. These results provide solutions for the practical design and performance evaluation of stellar interferometry.

\section{\label{sec:level2}Theoretical Model}
\begin{figure}[bt]
\includegraphics[width=0.40\textwidth]{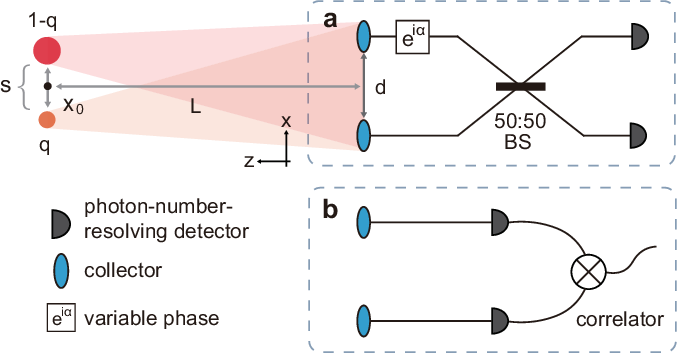}
\caption{\label{fig: schematic} Schematic of two-mode stellar interferometric systems, broadly categorized into (a) amplitude interferometry $G^{(1)}$, in which signals collected by two telescopes are coherently combined via an optical link, and (b) intensity interferometry $G^{(2)}$, in which signals are recorded independently and correlated digitally in post-processing. The key system parameters include the relative brightness $q$ of two thermal sources,  the average photon number $\bar{N}$ of each telescope mode, and the source-interferometer distance $L$. The spatial parameters to be estimated are the separation and the geometric position  between the sources $\boldsymbol{\theta}=(s,x_0)^{\top}$.}
\end{figure}

In the setup shown schematically in Fig.~\ref{fig: schematic}, radiation from two incoherent thermal point sources with unequal brightnesses is received by a two-mode interferometer operating in the far-field, paraxial regime. The quantum state received is a two-mode Gaussian state $\rho$, which is completely characterized by its first moment $\lambda_\mu=\mathrm{Tr}[\rho\mathbf{a}_\mu]$, with $\mathbf{a}=[a_1,a_1^\dagger,a_2,a_2^\dagger]$ and its second moment $\Sigma_{\mu\nu}=\frac12\operatorname{Tr}[\rho(\tilde{\mathbf{a}}_\mu\tilde{\mathbf{a}}_\nu+\tilde{\mathbf{a}}_\nu\tilde{\mathbf{a}}_\mu)]$, with $\tilde{\mathbf{a}}_\mu=\mathbf{a}_\mu-\lambda_\mu $. Here, $a_1,a_1^{\dagger}$ and $a_2,a_2^{\dagger}$ denote the annihilation and creation operators of the respective telescope modes.  The first and second moments, $\lambda_\mu$ and $\Sigma$, of state $\rho$ are given by \cite{pearce2017optimal,wang2021superresolution}
\begin{equation}
\label{cov}
\begin{aligned}
&\lambda_\mu=0~\forall\mu,\\
\Sigma&=\begin{pmatrix}0&c_1&0&b\\c_1&0&b^*&0\\0&b^*&0&c_2\\b&0&c_2&0\end{pmatrix},\\
\end{aligned}
\end{equation}
where $b=\langle\hat{a}_{2}^{\dagger}\hat{a}_{1}\rangle$, $c_{j}=\langle\hat{a}_{j}^{\dagger}\hat{a}_{j}\rangle+1/2$ and $\langle \cdots \rangle$ denotes the expectation value. We assume two telescope modes have the same average photon number as $\langle\hat{a}_{1}^{\dagger}\hat{a}_{1}\rangle=\langle\hat{a}_{2}^{\dagger}\hat{a}_{2}\rangle=\bar{N}$. The element $b$ and its complex conjugate $b^*$ are linked to the CDC between the two telescopes, which can be described as $\gamma=\langle\hat{a}_2^{\dagger}\hat{a}_1\rangle/{\left[\langle n_1\rangle\langle n_2\rangle\right]^{\frac12}}$ \cite{mandel1995optical}. Without loss of generality, we consider two incoherent point sources distributed along the $x$ direction, characterized by the intensity distribution
\begin{equation}
    I(x)=I_0[q\delta(x-x_1)+(1-q)\delta(x-x_2)],
\end{equation}
where $I_0$ is the total intensity of the sources, $x_{1,2}$ are the coordinates of two point sources, $q\in(0,\frac{1}{2}]$ is the relative brightness and $\delta(x)$ is the Dirac delta function. This model serves as a simplified representation of planetary systems, where a bright star's light is scattered by a dimmer planet. According to the Van Cittert–Zernike theorem  \cite{goodman2015statistical}, the far-field CDC is proportional to the Fourier transform of the intensity distribution $I(x)$ in the source plane, which gives
\begin{equation}
\label{gamma}
\begin{aligned}
   \gamma\propto&\int_{-\infty}^{\infty}I(x)e^{i\frac{kd}{L} x}dx=qe^{i\frac{kd}{L}x_1}+(1-q)e^{i\frac{kd}{L}x_2}, 
\end{aligned}
\end{equation}
where  $d$ is the length of the baseline, $k$ is the wave vector of the light and $L$ is the longitudinal distance to the source plane. The spatial properties can be recast into the separation between the point sources, given by $s=x_2-x_1$, and the geometric position between the sources, defined as $x_0=px_1+(1-p)x_2$, where $p$ depends on the chosen reference point. Certain choices of $p$ have clear physical significance, such as the geometric center ($p=\frac{1}{2}$), the intensity centroid ($p=q$), and the position of the brighter source ($p=0$). After the reparameterization, the amplitude and phase of the CDC in Eq.~(\ref{gamma}) can be expressed as follows
\begin{equation}
\label{gamma_phi}
 \vert \gamma \vert=\sqrt{1+2(q-1)q[1-\cos(\kappa s)]},~\arg(\gamma)=\kappa x_0+\phi_{p,s},
\end{equation}
where $\kappa=kd/L$ represents the effective aperture of the interferometer and $\phi_{p,s}$ is the phase term related to the values of $p$ and $s$.  The resolution of the imaging system is determined by the precision in estimating the separation  $s$.

\section{Multiparameter Quantum Fisher information}
\begin{figure}[t]
    \centering
    \includegraphics[width=0.47\textwidth]{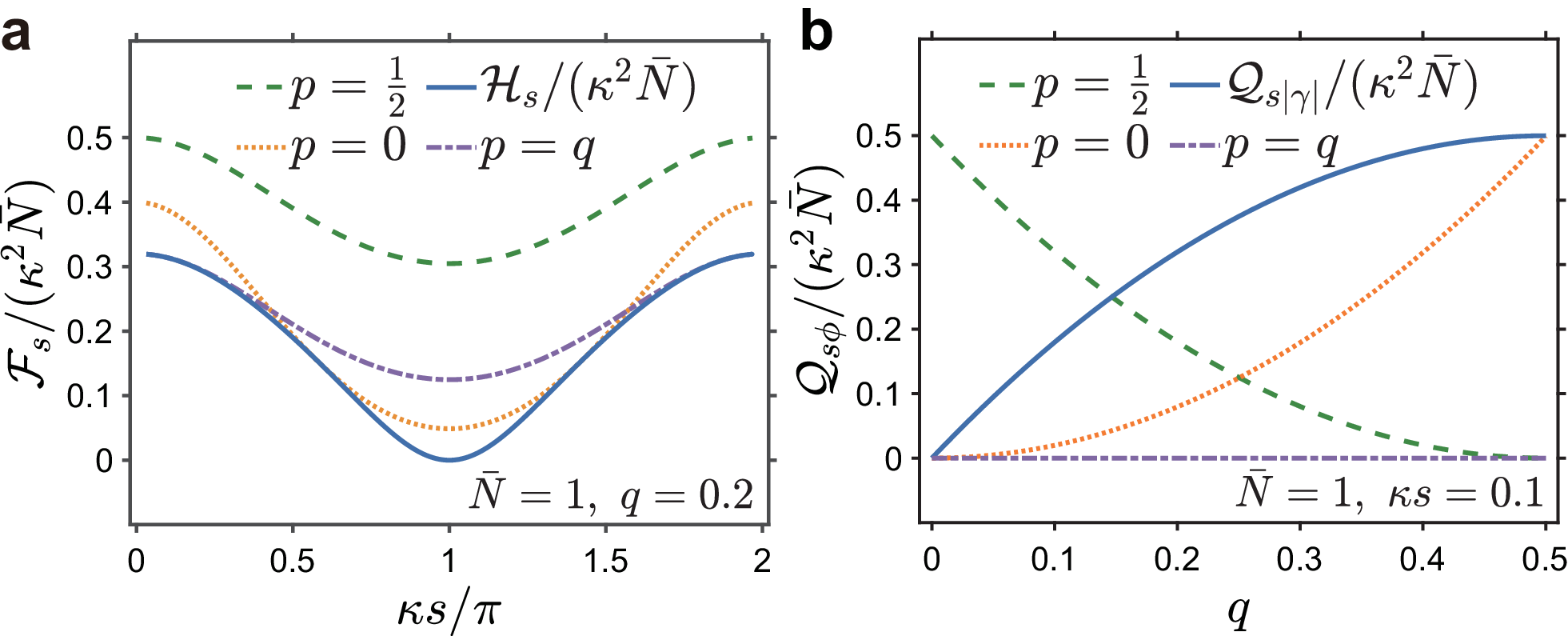}
     \caption{(a) The QFI for separation as a function of the normalized separation $\kappa s$ with $q=0.2$ and $\bar{N}=1$. (b) The QFI for separation contributed from the amplitude and phase of CDC as a function of different relative brightness $q$ with $\kappa s=0.1$ and $\bar{N}=1$.}
    \label{QFIs}
\end{figure}

In multiparameter quantum estimation problems, the central quantity is the QFI matrix. The quantum Cram\'{e}r-Rao bound states that the covariance matrix of the unbiased estimator of the parameters $\boldsymbol{\theta}$ is bounded by the inverse of the QFI matrix:
$
\Sigma_{{\boldsymbol{\theta}}}=\langle(\hat{\boldsymbol{\theta}}-\langle\hat{\boldsymbol{\theta}}\rangle)(\hat{\boldsymbol{\theta}}-\langle\hat{\boldsymbol{\theta}}\rangle)^{{\top}}\rangle\geq\mathcal{Q}^{-1}
$. The QFI is an intrinsic quantity of the quantum state whose closed form of a Gaussian state can be derived in terms of the first and second moments $\lambda_\mu$ and $\Sigma$ \cite{vsafranek2018estimation,gao2014bounds}. In our case,  $\lambda_\mu=0$, the expression simplifies to
 \begin{equation}
 \label{QFI}
\mathcal{Q}=\frac12\mathfrak{M}_{\alpha\beta,\mu\nu}^{-1}\left( \nabla_{\boldsymbol{\theta}} \Sigma_{\alpha\beta}\right)\left( \nabla_{\boldsymbol{\theta}} \Sigma_{\mu\nu}\right)^\top,
 \end{equation}
where $\mathfrak{M}=\Sigma\otimes\Sigma+\frac{1}{4}\Omega\otimes\Omega$ with $\Omega=\bigoplus_{k=1}^ni\sigma_y$, $\nabla_{\boldsymbol{\theta}}=(\frac{\partial}{\partial s},\frac{\partial}{\partial x_0},\frac{\partial}{\partial q})$ and the repeated indices imply summation. Notably, the QFI matrix is not of full rank for a two-mode state received in our interferometer, which limits the simultaneous measurement of all three parameters. In practice, the relative brightness $q$ can be predetermined through the eclipsing binaries, a phenomenon in which the photon flux of the binary system varies with the orbital motion of the companions \cite{prvsa2016physics,alonso2018characterization}. The main challenge lies in accurately estimating the spatial properties. Specifically, the separation $s$, reflecting the resolution of the imaging system, is of particular interest. If the geometric position $x_0$ for a given $p$ is known in advance, the precision of separation is determined by the reciprocal of the corresponding element of the QFI matrix, which is given by
\begin{equation}
\begin{aligned}
&\mathcal{Q}_s=\kappa^2\bar{N}\{(1-2p)^2[q^2-(q-1)q\cos(\kappa s)]+2p^2\\
&+(1-4p^2)q\}+\mathcal{O}\left[\bar{N}\right]^{2}\xlongequal{s\to0}2\kappa^2\bar{N}(p^{2}-2pq+q),
\end{aligned}
\label{Qs}
\end{equation}
where $\mathcal{O}\left[\cdot \right]^{n}$ denotes the contribution of higher-order terms ($\geq n$). This result extends the result in Ref.~\cite{wang2021superresolution}, which focused on symmetric source structures ($p = q = \frac{1}{2}$), to the case of an arbitrary relative brightness and reference point. In Fig.~\ref{QFIs}(a), $\mathcal{Q}_s$ is plotted for specific values of $p$ corresponding to different situations. $\mathcal{Q}_s$ exhibits periodic dependence on separation, with a pronounced decline at intermediate values within the period. When the geometric center $(p=1/2)$ is known, this decline arises from the net result of multiphoton contributions \cite{nair2016far}, contrasting with the weak-source limit where $\mathcal{Q}_s$ remains separation-independent \cite{fiderer2021general,lupo2020quantum}. For more practical scenarios where either the intensity centroid $(p=q)$ or the position of the brighter source $(p=0)$ is known, $\mathcal{Q}_s$ becomes sensitive to the relative brightness $q$ and decays rapidly throughout the period. This reduction is driven not only by thermal effects but also by the choice of reference point.

If the geometric position $x_0$ is unknown, joint estimation of both parameters $s$ and $x_0$ is required. A more fundamental precision for separation estimation can be obtained by taking the reciprocal of the corresponding element of the inverse QFI matrix, which yields
\begin{equation}
\begin{aligned}
\mathcal{H}_{s}=\frac1{[\mathcal{Q}^{-1}]_{11}}&=\frac{\kappa^2\bar{N}(1-q)q[(]1+\cos(\kappa s)]}{1+2(q-1)q[(]1-\cos(\kappa s)]}+\mathcal{O}\left[\bar{N}\right]^{2}\\
  &\xlongequal{s\to0}2\kappa^2\bar{N}(1-q)q.   
\end{aligned}
\label{Hs}
\end{equation}
$\mathcal{H}_s$ still remains constant as the separation approaches zero, demonstrating the potential of interferometry to achieve superresolution. Moreover, $\mathcal{H}_s$ establishes a lower bound for $\mathcal{Q}_s$, as shown in Fig.~\ref{QFIs}(a). This is reasonable since the geometric position $x_0$ serves as a nuisance parameter in this context \cite{suzuki2020quantum}; a known $x_0$ provides additional information that can enhance the precision of separation estimation. The difference between Eqs.~(\ref{Qs}) and~(\ref{Hs}) highlights the impact of the nuisance parameter on the desired separation estimation, which can be further elucidated through general error analysis. Specifically, we analyze the precision of separation via error propagation from the QFI for the amplitude and phase of CDC measurements (detailed in Appendix~\ref{CDC}). The precision of separation satisfies the following inequality
\begin{equation}
 \mathrm{var}(s)\geq\left\{\begin{array}{ll}\mathcal{Q}^{-1}_s=({\mathcal{Q}_{s\vert \gamma \vert} + \mathcal{Q}_{s\phi}})^{-1}&(x_0\text{ is known})\\\\\mathcal{H}^{-1}_s=(\mathcal{Q}_{s|\gamma|})^{-1}&(x_0\text{ is unknown})\end{array}\right.   
 \label{inequality}
\end{equation}
where $\mathrm{var}(s)$ is the variance of the separation estimation, $\mathcal{Q}_{s|\gamma|}=(\partial |\gamma|/\partial s)^2\mathcal{Q}_{|\gamma|}$, and $\mathcal{Q}_{s\phi}=(\partial \phi/\partial s)^2\mathcal{Q}_{\phi}$, with $\mathcal{Q}_{|\gamma|}$ and $\mathcal{Q}_{\phi}$ denoting the QFI of estimating the amplitude and phase of CDC, respectively. The full information about the separation can be divided into two parts: the amplitude information $\mathcal{Q}_{s|\gamma|}$ and the phase information $\mathcal{Q}_{s\phi}$. Notably, amplitude information $\mathcal{Q}_{s|\gamma|}$ proves to be more robust for separation estimation, as phase information $\mathcal{Q}_{s\phi}$ is diminished by the presence of the nuisance parameter. We plot $\mathcal{Q}_{s|\gamma|}$ and $\mathcal{Q}_{s\phi}$ with specific values of $p$ for different relative brightnesses in Fig.~\ref{QFIs}(b). For certain geometric priors, such as a fixed geometric center ($p=\frac12$), phase information $\mathcal{Q}_{s\phi}$ becomes more valuable at low relative brightness $q$. However, for the experimentally more relevant scenarios in which only the intensity centroid ($p=q$) can be reliably determined, $\mathcal{Q}_{s\phi}$ approaches zero for any relative brightness. Under such conditions, accessing the phase information of the CDC does not  directly contribute to improved resolution.

\section{Fisher information of interferometric schemes}
In classical parameter estimation, the covariance matrix of unbiased parameter estimates is bounded by the inverse of the Fisher information (FI) matrix. The QFI provides the maximum achievable FI, which states that the covariance matrix follows the chain inequality: $\Sigma_{{\boldsymbol{\theta}}} \geq \mathcal{F}^{-1} \geq \mathcal{Q}^{-1}$. For two-mode interferometry, the FI matrix can be defined as
\begin{equation} \mathcal{F}=\sum_{m_1}\sum_{m_2}\frac{[\nabla_{\boldsymbol{\theta}} p^{(i)}(m_1,m_2)][\nabla_{\boldsymbol{\theta}} p^{(i)}(m_1,m_2)]^\top}{p^{(i)}(m_1,m_2)}, 
\label{FI_caculation}
\end{equation} 
where $p^{(i)}(m_1,m_2)$ is the conditional probability distribution of observing an outcome $(m_1,m_2)$ under the interferometric scheme $G^{(i)}$.

For schemes based on amplitude interferometry, known as $G^{(1)}$ schemes, a general apparatus is shown in Fig.~\ref{fig: schematic}(a). In this setup, light collected by two telescopes is coherently combined on a beam splitter (BS), with a variable phase delay $\alpha$ introduced in one arm, followed by photon-number-resolved detection at both output ports. The action of amplitude interferometry can be described by a unitary operation $d_{1}=\frac{1}{\sqrt2}(a_{1}+e^{i\alpha}a_{2})$ and $d_{2}=\frac{1}{\sqrt2}(a_{1}-e^{i\alpha}a_{2})$, where $d_{1,2}$ are the annihilation operators of output modes. The measurement corresponds to projecting the output state onto the Fock basis $\vert m_1,m_2\rangle_d$ of modes $d_{1,2}$. 

For schemes based on intensity interferometry, known as $G^{(2)}$ schemes, the apparatus is illustrated in Fig.~\ref{fig: schematic}(b). Each mode of the state is detected separately, corresponding to a direct projection onto the Fock basis $\vert m_1,m_2\rangle_a$ of modes $a_{1,2}$. The analytical expressions for the conditional probability distributions $p^{(1)}(m_1,m_2)$ and $p^{(2)}(m_1,m_2)$, corresponding to amplitude interferometry and intensity interferometry, are provided in Appendix~\ref{probability}. In the following, we compare both interferometric schemes by calculating the FI for separation, assuming prior knowledge of the geometric position $x_0$ and relative brightness $q$. 

\subsection{Amplitude interferometry}

\begin{figure}[tbp]
    \centering
    \includegraphics[width=0.47\textwidth]{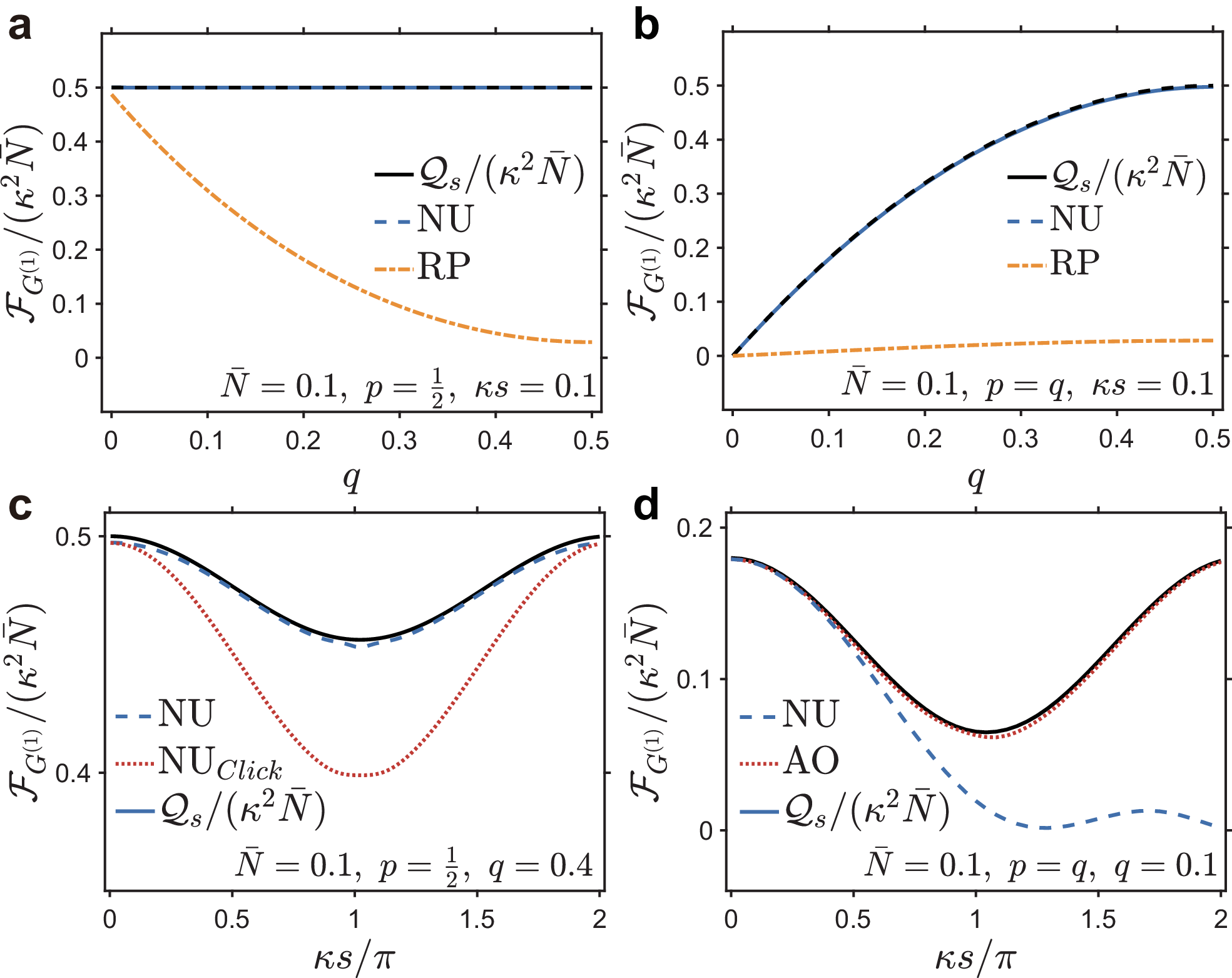}
  \caption{Performance of amplitude interferometry-based strategies. (a) and (b) The FI for separation as a function of relative brightness $q$ with a known  geometric center and intensity centroid for a small separation. (c) and (d) The FI for separation as a function of normalized separation $\kappa s$ with a known geometric center and  intensity centroid. The interferometric strategies evaluated include adaptive-optimal (AO), nulling (NU), and random-phase (RP) strategies. The average photon number is $\bar{N} = 0.1$, and the interferometer is aligned to the known geometric position with $\kappa x_0 = 0$.}
    \label{2}
\end{figure}

In the case of two identical thermal sources with a known geometric center ($p = q = \frac{1}{2}$), recent work has shown that amplitude interferometry can achieve the QFI for separation \cite{wang2021superresolution}.  For unequal brightnesses and alternative reference points, we provide analytical and numerical evidence that amplitude interferometry can still attain the QFI for separation across all source parameter values. Notably, these findings remain applicable in the single-photon regime, as elaborated in Appendix~\ref{single-photon regime}.

The maximum FI for separation is achieved by selecting an optimal phase delay, denoted as $\alpha = \alpha_\mathrm{op}$. Given that the optimal phase delay $\alpha_\mathrm{op}$ may depend on the source parameters, it is essential to adaptively adjust $\alpha$ to attain optimal precision. This process is referred to as the adaptive-optimal (AO) strategy. To explore a parameter-independent alternative, we investigate the nulling (NU) strategy, initially proposed in astronomy for exoplanet detection \cite{glindemann2011principles,david2015practical}. The basic idea behind the NU strategy is to suppress radiation from the brighter source through destructive interference. In this context, the NU strategy corresponds to fixing the phase delay at either $\alpha = 0$ or $\pi$ \cite{labeyrie2006introduction}. Considering all detection outcomes with $p^{(1)}(m_1,m_2)$ for {$m_1+m_2\leq7$}, the corresponding FI for separation with NU strategy is given by
\begin{equation}
\begin{aligned}
   \mathcal{F}_{G^{(1)}}|&_{\mathrm{NU}}=\kappa^2\bar{N}\{2(p^2-2pq+q)-\frac{\kappa^2(1-2p)^2(1-q)q}{2}s^2\\
&+\mathcal{O}\left[s\right]^{3}\}+\mathcal{O}\left[\bar{N}\right]^{2}\xlongequal{s\to0}2\kappa^2\bar{N}(p^{2}-2pq+q),  
\end{aligned}
\label{NU}
\end{equation}
where terms beyond the second order in $s$ are omitted for brevity. The leading term of $\mathcal{F}_{G^{(1)}}|_{\mathrm{NU}}$ aligns with that of $\mathcal{Q}_s$ in Eq.~(\ref{Qs}). Furthermore, the absence of first-order terms in $s$ ensures that $\mathcal{F}_{G^{(1)}}|_{\mathrm{NU}}$ equals $\mathcal{Q}_s$ for all relative brightness values when the  separation is small ($\kappa s < 1$), as shown in Figs.~\ref{2}(a) and \ref{2}(b). Further analysis in Appendix~\ref{probability} reveals that the destructive interference outcomes contribute to FI values approaching  $\mathcal{Q}_s$.

For larger separations ($\kappa s > 1$), source asymmetries can lead to deviations from quantum-optimal performance when employing the NU strategy. When the geometric center is known ($p = \frac{1}{2}$), the sources remain symmetric about the reference point, and $\mathcal{F}_{G^{(1)}}|_{\mathrm{NU}}$ matches $\mathcal{Q}_s$ for any separations, even in cases of unequal brightnesses [Fig.~\ref{2}(c)]. However, for asymmetric sources in both brightness and reference points (i.e., $p \neq \frac{1}{2},~q \neq \frac{1}{2}$), $\mathcal{F}_{G^{(1)}}|_{\mathrm{NU}}$ no longer achieves $\mathcal{Q}s$ for large separations, as depicted in Fig.~\ref{2}(d). In such instances, the AO strategy is necessary to sustain quantum-optimal performance. 

In practice, detectors, such as avalanche photodiodes, can distinguish only the presence or absence of photons. For these systems, the FI without photon-number-resolving detection is denoted as "click". As illustrated in Figs.~\ref{2}(c), applying the same strategies with click detection retains a significant portion of the original FI. Superresolution is also preserved, as indicated by the achievement of $\mathcal{Q}_s$ in the limit of vanishing separation. Additionally, phase fluctuation can degrade the performance of both the AO and NU strategies. Instead of stabilizing the optimal phase delay $\alpha_\mathrm{op}$ through feedback mechanisms, the random-phase (RP) strategy provides a simpler alternative, where $\alpha$ is randomly sampled over $[0, 2\pi)$. Notably, the phase delay $\alpha$ can be monitored during the measurement. While the RP strategy does not achieve quantum-optimal performance for arbitrary source parameters, it remains effective for small separations and extreme brightness contrasts, as shown in Figs. \ref{2}(a) and \ref{2}(b). These conditions are particularly relevant for exoplanet detection in which traditional imaging techniques are generally inadequate.

\subsection{Intensity interferometry}
Intensity interferometry measures the intensity correlation between different telescopes to observe the second-order coherence of light. The intensity correlation is derived from coincidence events between the detectors, while such events would be neglected if the quantum states were truncated to the single-photon level \cite{fiderer2021general,lupo2020quantum}. For weak thermal strength ($\bar{N}\ll1$), the intensity correlation is primarily characterized by the two-photon coincidence events. The probability distribution can be expressed as
\begin{equation}
\begin{aligned}
  p^{(2)}(1,1)&=\frac{\bar{N}^2\left[1+\vert \gamma \vert^2+\bar{N}(1-\vert \gamma \vert^2)\right]}{\left[1+2\bar{N}+\bar{N}^2(1-\vert \gamma \vert^2)\right]^3}\\
  &=\bar{N}^2(1+\vert \gamma \vert^2)+\mathcal{O}\left[\bar{N}\right]^{3},  
\end{aligned}
\end{equation}
where $\vert \gamma \vert$ is defined in Eq. (\ref{gamma_phi}). This probability distribution contains no phase information and is proportional to the normalized second-order coherence function via the Siegert relation $g^{(2)}=1+\vert \gamma \vert^2$ \cite{drechsler2022revisiting}.  In contrast, amplitude interferometry measures the first-order coherence of light, which provides both the amplitude and phase information of the CDC.

By considering all detection outcomes with $p^{(2)}(m_1,m_2)$ for $m_1+m_2\leq7$, the FI for separation using intensity interferometry is given by
\begin{equation}
\mathcal{F}_{G^{(2)}}= \frac{2\kappa^2\bar{N}^2(q-1)^2q^2\sin(\kappa s)^2}{1+(q-1)q[(]1-\cos(\kappa s)]}+\mathcal{O}[\bar{N}]^3\xlongequal{s\to0}0.  
\end{equation}
In comparison to the FI for separation in amplitude interferometry, $\mathcal{F}_{G^{(2)}}$ diminishes as the separation tends to zero. Consequently, superresolution is unattainable for intensity interferometry, as the diminishing FI leads to infinite uncertainty in the separation estimation. It is widely recognized that the limitations of intensity interferometry primarily arise from the low probability of the post-selection of coincidence events.  Our analysis reveals that the two-photon coincidence event $p^{(2)}(1,1)$ contributes only a minor amount to the FI for separation. Furthermore, the dominant events, $p^{(2)}(0,1)$ and $p^{(2)}(1,0)$, contribute even less significantly (see Appendix~\ref{probability}). In contrast to amplitude interferometry, the distinct behavior of intensity interferometry results in reduced precision for separation estimation, thereby precluding the achievement of superresolution. We also evaluate the performance of heterodyne detection, as detailed in Appendix~\ref{heterodyne}. This scheme, similar to intensity interferometry, also fails to achieve superresolution.

\section{quantitative comparison between the two interferometric schemes}
It is indisputable that amplitude interferometry outperforms intensity interferometry under ideal conditions, assuming identical baseline length and average photon numbers. However, amplitude interferometry is vulnerable to optical imperfections and transmission loss. While such effects are well recognized in interferometric systems \cite{kurdzialek2023measurement, demkowicz2009quantum}, that impact on resolution has not been systematically examined in the context of interferometric imaging. In contrast, intensity interferometry is largely insensitive to these practical limitations. This insensitivity allows intensity interferometry to utilize telescopes with diameters of several meters, positioned over distances of several kilometers, thereby increasing the average photon number and enlarging the baseline length.  Recent work \cite{bojer2022quantitative} showed that intensity interferometry may offer performance advantages under such practical conditions.  However, their analysis does not fully capture the characteristics of distributed telescope architectures or the thermal statistics. To enable a more rigorous and metrologically grounded comparison, we refine the comparison to address the effects of measurement imperfections, optical losses, baseline length, and photon number.  For a fair comparison, we assume the intensity centroid ($p=q$) is known, whereupon the contribution of phase information to the resolution becomes negligible, eliminating any bias due to the inability of intensity interferometry to access the CDC phase.

The primary measurement imperfection in amplitude interferometry stems from cumulative decoherence due to optical imperfections and atmospheric turbulence, which manifest as reduced visibility $\mathcal{V}$ and deviation from the optimal phase delay, $\delta = \alpha - \alpha_\mathrm{op}$.  We assume that both reduced visibility $\mathcal{V}$ and phase deviation $\delta$ can be monitored using a reference, which is well-established in modern interferometers \cite{glindemann2011principles}. The derivation of the modified conditional probability distributions $\tilde{p}^{(i)}(m_1, m_2)$ is provided in Appendix~\ref{relation}. As shown in Fig.~\ref{3}, the FI for separation using amplitude interferometry achieves the QFI for separation with perfect visibility $\mathcal{V} = 1$ and zero phase deviation $\delta = 0$. However, even a small reduction in visibility $\mathcal{V}$ progressively degrades the performance of amplitude interferometry. In contrast, small phase deviations have a slight impact on its quantum-optimal performance. The FI for amplitude interferometry converges to that of intensity interferometry when $\mathcal{V} = 0$ or $\delta = \pi/2$, highlighting that intensity interferometry is a special case of amplitude interferometry without first-order interference. This explains its inefficiency in extracting information from the CDC and its inability to achieve superresolution, as it operates in a setup fundamentally opposite to the optimal strategy of amplitude interferometry.

\begin{figure}[t]
    \centering
    \includegraphics[width=0.47\textwidth]{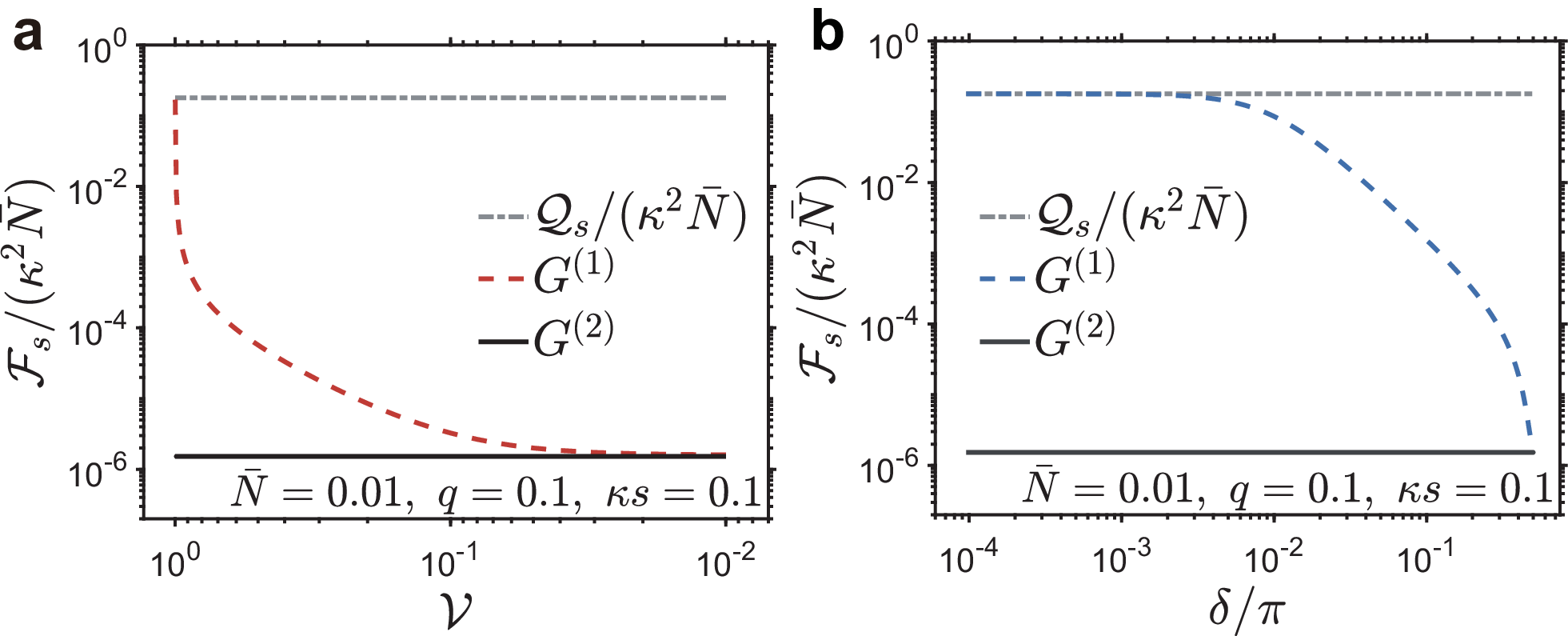}
    \caption{Comparison of interferometric schemes in the lossless case with a known intensity centroid. (a) The FI for separation as a function of reduced visibility $\mathcal{V}$. (b) The FI for separation as a function of phase deviation $\delta$. The parameters are set as $p = q = 0.1$, $\kappa s = 0.1$, and $\bar{N} = 0.01$. For these values, the optimal phase delay in amplitude interferometry is $\alpha_\mathrm{op} = 0$.}
    \label{3}
\end{figure}

\begin{figure}[b]
    \centering
    \includegraphics[width=0.47\textwidth]{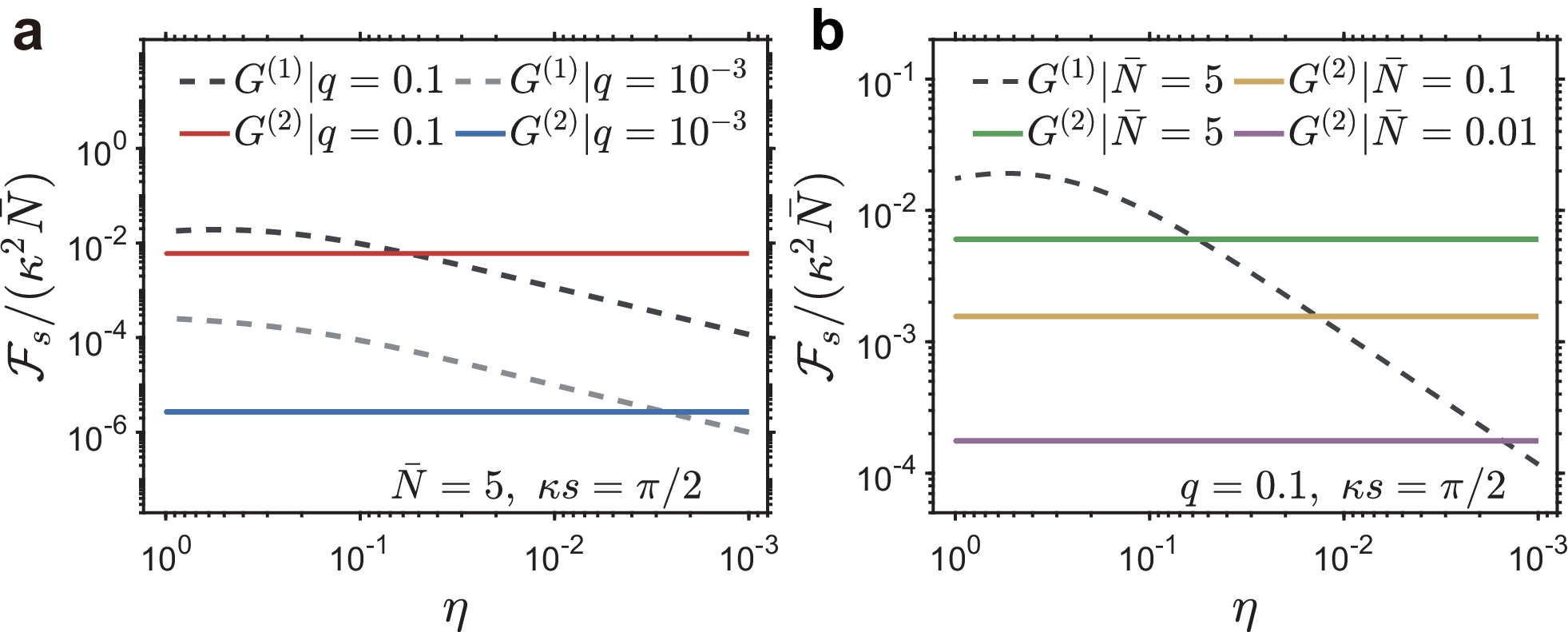}
    \caption{Comparison of lossy interferometric schemes with a known intensity centroid. (a) The FI for separation as a function of transmission  parameter $\eta$ for different relative brightnesses $q$ with $\kappa s=\pi/2$ and $\bar{N}=5$. (b) The FI for separation as a function of $\eta$ for different average photon numbers $\bar{N}$ with  $\kappa s=\pi/2$ and $q=0.1$. The phase delay in amplitude interferometry is fixed at $\alpha = 0$.}
    \label{4}
\end{figure}

Optical loss also affects amplitude interferometry by reducing the average photon number $\bar{N}$ \cite{marchese2023large,huang2024limited}, effectively replacing it with $\eta\bar{N}$, where $\eta$ is the transmission parameter. In contrast, no optical loss is incurred in intensity interferometry, since only electrical signals are connected between telescopes.  A comparison between lossy amplitude interferometry and intensity interferometry for various relative brightness levels and average photon numbers is presented in Fig.~\ref{4}. For moderate  transmission loss ($\eta\bar{N}>0.1$), increasing loss has a limited impact on the FI of amplitude interferometry. This is due to the fact that a reduction in the effective photon number $\eta\bar{N}$ suppresses the detrimental net effect of multiphoton events, thereby partially compensating the loss-induced degradation. In contrast, under large losses and in the weak thermal source regime, the FI of amplitude interferometry decreases approximately linearly with $\eta$, making its advantage no longer ensured. As a result, when loss is sufficiently high, intensity interferometry can outperform amplitude interferometry in precision. Nevertheless, since the FI for intensity interferometry scales as $\bar{N}^2$ and $q^4$, while for amplitude interferometry it scales as $\bar{N}$ and $q^2$, amplitude interferometry outperforms for  weak thermal sources ($\bar{N}<0.1$) and high-contrast targets ($q \ll 0.1$), even with significant loss.

Considering these practical limitations, further investigation is required to determine the optimal trade-off between the sensitivity of amplitude interferometry and the robustness of intensity interferometry under specific observation conditions and source characteristics. Thus, we compare several state-of-the-art stellar interferometers in the benchmark task of resolving two adjacent stars of planetary systems. Since the NU strategy is quantum optimal for a small separation, we select the Keck Interferometer Nuller (KIN) and the Large Binocular Telescope Interferometer (LBTI) as representative nulling amplitude interferometers. KIN, the first mid-infrared nulling interferometer, employs 10-m telescopes with a maximum baseline of 85 m \cite{serabyn2012keck}. The LBTI, on the other hand, features two 8.4-m primary mirrors with a 14.4-m baseline \cite{defrere2016nulling}. As for the intensity interferometer, the most prominent candidate is the Very Energetic Radiation Imaging Telescope Array (VERITAS), utilizing four 12-m telescopes with a maximum baseline of 172.5 m \cite{abeysekara2020demonstration}. Although the current intensity interferometer relies on relatively modest baselines, the prospects for ultra-long baselines are promising. The Cherenkov Telescope Array (CTA), for instance, envisions 50 to 100 telescopes with apertures ranging from 5 to 25 m and baselines up to 2 km \cite{dravins2013optical}. Based on this, we considered a configuration of $n=100$ telescopes with 15 m apertures in the CTA. Since the electrical signal can be freely copied and combined, it becomes feasible to synthesize an arbitrary number of baselines between any pair of telescopes on- or offline, with $n$ telescopes yielding up to $n(n-1)/2$ baselines. To illustrate the advantages of multiple baselines, we examined two schemes: $\mathrm{CTA}_{100\mathrm{Tele}}$, utilizing all distributed telescopes, and $\mathrm{CTA}_{2\mathrm{Tele}}$, employing only the two telescopes at the maximum separation. Our simulations accounted for the telescope sizes and baselines of these interferometers, as summarized in Table~\ref{table1}.  The assumed astronomical target is a planetary  system with a relative brightness of $q=0.1$, located 10 pc away ($L\approx 3 \times 10^{16}~\mathrm{m}$). The photon flux is $\bar{N}=10^{-6}/\mathrm{m}^2$ per coherence time at a wavelength of  $1 \mu$m. The visibility of amplitude interferometry is set to $\mathcal{V}=0.99$ with phase fluctuation excluded.  It is important to note that we employed an unrealistically small angular distance to illustrate the interferometer's precision in the sub-Rayleigh region.
\begin{table}[t]
\centering
\begin{tabular}{|c|c|c|c|c|}\hline
Method&Interferometer & Baseline  & Diameter (m)&$n_{\mathrm{Tele}}$\\\hline
\multirow{2}{*}{$G^{(1)}$}&Keck &$85$ m&$10$ &2\\
\multirow{2}{*}{}&LBTI &$14.4$ m &$8.4$  &2\\\hline
\multirow{3}{*}{$G^{(2)}$}&VERITAS &$172.5$ m&$12$  &2\\
\multirow{3}{*}{}&CTA$_{2\mathrm{Tele}}$ &$2$  km &$15$ &2 \\
\multirow{3}{*}{}&CTA$_{100\mathrm{Tele}}$ &$2$ km&$15$  &100 \\\hline
\end{tabular}
\caption{The different baselines, telescope sizes, and numbers of telescopes of five stellar interferometers used for the numerical calculations in Fig. \ref{g1g2compare}.}
\label{table1}
\end{table}

\begin{figure}[pbth]
    \centering
    \includegraphics[width=0.43\textwidth]{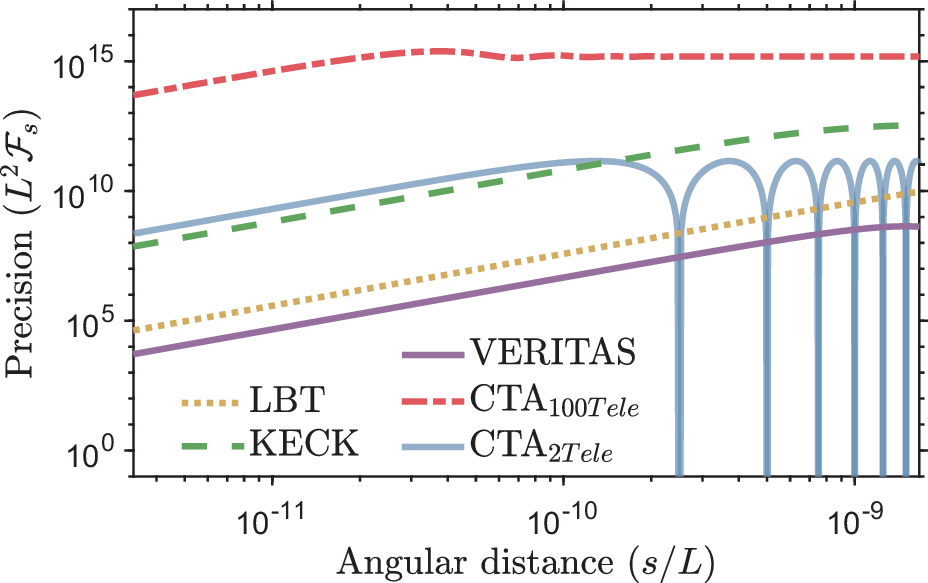}
    \caption{The FI for separation with five stellar interferometers investigated in this paper against the angular separation under realistic exoplanet-detection conditions.}
    \label{g1g2compare}
\end{figure}

The theoretical precision limits are presented in Fig.~\ref{g1g2compare}, which highlights the competitive performance of large-baseline intensity interferometers relative to amplitude interferometry. As precision scales quadratically with the effective aperture $\kappa$, extending the baseline significantly enhances the performance of intensity interferometry. However, rapid oscillations occur due to its periodic dependence on the normalized separation $\kappa s$, as demonstrated by the performance of $\mathrm{CTA}_{2\mathrm{Tele}}$. This challenge can be mitigated through the use of multiple baselines, as exemplified by $\mathrm{CTA}_{100\mathrm{Tele}}$, where varying values of $\kappa s$ are captured by adjusting $\kappa$. This approach also alleviates ambiguities in separation estimation. By electronically replicating the digital signals from each telescope, synthesizing multiple baselines effectively simulates multiple experiments within a limited observation window, leading to precision that surpasses even amplitude interferometry. These results show the potential that most astronomical observations achievable with the amplitude interferometer can be effectively verified by the intensity interferometer with enhanced resolution utilizing presumably much simpler setups. 

\section{Conclusion}
In summary, we employed the quantum estimation theory to determine the ultimate limit and achievable precision of stellar interferometry for the benchmark problem of resolving two unequal-brightness thermal sources. In a multiparameter estimation framework, we analytically derived the ultimate limit for separation estimation by treating geometric position (i.e., the geometric center, intensity centroid, and brighter source position) as a nuisance parameter. Our analysis revealed that the QFI for separation can be separated into contributions from the amplitude and phase of the CDC, while the uncertainty in geometric position wipes out any benefit from measuring the phase information.  We demonstrate that amplitude interferometry can approach the QFI for separation and achieve superresolution. In contrast, intensity interferometry is highly inefficient in separation estimation, which fails to maintain constant precision in the limit of vanishing separation. 

For practical considerations, we quantitatively compare two interferometric schemes in terms of measurement imperfections and optical loss. While intensity interferometry offers no metrological advantage in the lossless case, it outperforms amplitude interferometry when significant optical loss is considered. Furthermore, the potential to employ ultra-long baselines with multiple telescopes makes intensity interferometry a highly competitive technique for real-world astronomical applications.  Despite the challenges posed by the faintness of planetary signals and the stringent requirements of stellar interferometry, our work underscores the growing potential of large-scale interferometry for high-precision astronomical observations.

\begin{acknowledgements}
This work was supported by Innovation Program for Quantum Science and Technology (grant no. 2024ZD0300900), National Natural Science Foundation of China (grant nos. 12347104, U24A2017, and 12461160276), the National Key Research and Development Program of China (Grants
No.2023YFC2205802), Natural Science Foundation of Jiangsu Province (grant nos. BK20243060 and BK20233001), China Postdoctoral Science Foundation
(No. 2022M721565), in part by State Key Laboratory of Advanced Optical Communication Systems and Networks, China.
\end{acknowledgements}

\appendix

\section{The QFI matrix of two-mode interferometry}

The  QFI matrix for estimating $\boldsymbol{\theta}=(s,x_0,q)^{\top}$ takes the form
\begin{equation}
\label{QFIM}
   \mathcal{Q}_{\boldsymbol{\theta}}= \left(\begin{array}{ccc}\mathcal{Q}_{s}&\mathcal{Q}_{sx_0}&\mathcal{Q}_{s q}\\\mathcal{Q}_{sx_0}&\mathcal{Q}_{x_0}&\mathcal{Q}_{x_0q}\\\mathcal{Q}_{sq}&\mathcal{Q}_{x_0q}&\mathcal{Q}_{q}\end{array}\right),
\end{equation}
where the remaining diagonal elements are given by
\begin{equation}
\begin{aligned}
\label{A2}
 &\mathcal{Q}_{x_0}=2\kappa^2\bar{N}\{1+2(q-1)q[1-\cos(\kappa s)]\}+\mathcal{O}\left[\bar{N}\right]^{2},\\
 &\mathcal{Q}_{q}=\frac{\bar{N}[\cos(\kappa s)-1]}{(q-1)q}+\mathcal{O}\left[\bar{N}\right]^{2}\xlongequal{s\to0}0.\\
 \end{aligned}
\end{equation}
The off-diagonal terms are
\begin{equation}
\begin{aligned}
&\mathcal{Q}_{sx_0}=2\kappa^2\bar{N}[\{p-q^2+(q-1)q\\
&\times (2p-(2p-1)\cos(\kappa s))\}+\mathrm{O}\left[\bar{N}\right]^{2};\\
&\mathcal{Q}_{sq}=\kappa\bar{N}(1-2p)\sin(\kappa s)+\mathrm{O}\left[\bar{N}\right]^{2};\\
&\mathcal{Q}_{x_0q}=-2\kappa\bar{N}\sin(\kappa s)+\mathrm{O}\left[\bar{N}\right]^{2}.\\
\end{aligned}
\end{equation} 

The QFI matrix, without taking the first approximation in Eq.~(\ref{QFIM}), is of rank 2, implying that at most two parameters can be simultaneously estimated using a two-mode interferometer. Given the vanishing behavior of $\mathcal{Q}_q$ [Eq. (\ref{A2})] in the Rayleigh limit, interferometry proves inefficient for estimating relative brightness. We therefore restrict our analysis to the estimation of spatial parameters $\boldsymbol{\theta}=(s, x_0)^{\top}$, which are more amenable to interferometric techniques. To estimate additional parameters simultaneously, the mode of interferometry must be increased. The extended analysis of  multi-mode interferometry can be achieved by expanding the covariance matrix in Eq.~(\ref{cov}) to describe multi-mode Gaussian states, as shown in Ref.~\cite{wang2021superresolution}.

\section{General error analysis of parameter estimation}
\label{CDC}
The precision of the separation can be analyzed by the error propagation from the precision of the amplitude and phase of the CDC measurement. The QFI matrix for estimating  $\boldsymbol{\zeta}=(|\gamma|,\phi)^T$ was well established in Ref.~\cite{pearce2017optimal}, given by 
\begin{equation}
\label{QFI_gamma_phi}
 \mathcal{Q}_{\boldsymbol{\zeta}}=\left(\begin{array}{cc}\frac{2\bar{N}(1+\bar{N}+\bar{N}\gamma^2)}{\left(1-\gamma^2\right)\left[\bar{N}^2\left(1-\gamma^2\right)+2\bar{N}+1\right]}&0\\0&\frac{2\bar{N}\gamma^2}{\bar{N}(1-\gamma^2)+1}\end{array}\right).
\end{equation}

When the geometric position $x_0$ is known, the separation can be represented as a function of the amplitude and phase of CDC, and the estimator can be constructed as 
\begin{equation}
\label{estimator}
    \hat{s}=\alpha_{|\gamma|} \hat{s}_{|\gamma|}+\alpha_{\phi} \hat{s}_{\phi},
\end{equation}
where $\alpha_{|\gamma|,\phi}$ denotes the normalization coefficient with $\alpha_{|\gamma|}+\alpha_\phi=1$. By using standard error analysis, the minimum mean square error (MSE) can be written as
\begin{equation}
\begin{aligned}
      \mathrm{var}(s)&=\sum_{\zeta=|\gamma|,\phi}\alpha_{\zeta}^2(\frac{\partial s}{\partial\zeta})^2\mathrm{var}(\zeta)\geq\sum_{\zeta=|\gamma|,\phi}\alpha_{\zeta}^2(\frac{\partial s}{\partial\zeta})^2\frac{1}{\mathcal{Q}_\zeta},
\end{aligned}
\label{error_propagation}
\end{equation}
where $\mathrm{var}(\cdot)$  represents the variance of the estimated parameter, the $\mathcal{Q}_{|\gamma|,\phi}$ denote the QFI for estimating the amplitude and phase of the CDC. To calculate the optimal precision, we need to allocate these normalization coefficients to obtain the minimum value of the MSE. Equation ~(\ref{error_propagation}) is a multivariate quadratic function, and its minimum value can be calculated using the Hessian matrix \cite{binmore2002calculus}. The optimal coefficient is 
\begin{equation}
\label{coefficient_s}
\alpha_\zeta=(\frac{\partial\zeta}{\partial s})^2\mathcal{Q}_{\zeta}/\sum_{\zeta=|\gamma|,\phi}(\frac{\partial\zeta}{\partial s})^2\mathcal{Q}_{\zeta}=\mathcal{Q}_{s\zeta}/\sum_{\zeta=|\gamma|,\phi}\mathcal{Q}_{s\zeta},
\end{equation}   
where $\mathcal{Q}_{s|\gamma|}=(\frac{\partial |\gamma|}{\partial s})^2\mathcal{Q}|\gamma|$ and $\mathcal{Q}_{s\phi}=(\frac{\partial \phi}{\partial s})^2\mathcal{Q}\phi$ denote the QFI for separation contributed by the amplitude and phase of the CDC. Substituting Eqs.~(\ref{QFI_gamma_phi}) and~(\ref{coefficient_s}) into Eq.~(\ref{error_propagation}), we find the minimum MSE of separation is given exactly by the inverse of the QFI of the separation, which is  
\begin{equation}
\mathrm{var}(s)\geq\frac{1}{\mathcal{Q}_{s|\gamma|}+\mathcal{Q}_{s\phi}}=\frac{1}{\mathcal{Q}_s}.
\end{equation}

When the geometric position $x_0$ is unknown,  the precision bound for jointly estimating $\boldsymbol{\theta}=(s,x_0)^T$ is given by \cite{kay1993fundamentals}
\begin{equation}
   \Sigma_{{\boldsymbol{\theta}}} \geq \mathcal{F}_{\boldsymbol{\theta}}^{-1}= \mathcal{J}_{\boldsymbol{\theta}}\mathcal{F}_{\boldsymbol{\zeta}}^{-1} \mathcal{J}_{\boldsymbol{\theta}}^{\dagger}\geq \mathcal{J}_{\boldsymbol{\theta}}\mathcal{Q}_{\boldsymbol{\zeta}}^{-1} \mathcal{J}_{\boldsymbol{\theta}}^{\dagger}, 
\end{equation}
where  $\mathcal{J}_{\boldsymbol{\theta}}=\partial\boldsymbol{\theta}/\partial\boldsymbol{\zeta}$ is the Jacobian matrix with $2\times2$ dimensions. The Jacobian matrix can be derived from Eq.~(\ref{gamma_phi}), given by
\begin{equation}
    \mathcal{J}_{\boldsymbol{\theta}}=\left(\begin{array}{cc}\frac{\partial s}{\partial |\gamma|}&0\\\frac{\partial x_0}{\partial |\gamma|}&\frac{\partial x_0}{\partial \phi}\end{array}\right)=\left(\begin{array}{cc}\frac{\partial s}{\partial |\gamma|}&0\\\frac{\partial x_0}{\partial \phi}\frac{\partial \phi}{\partial s}\frac{\partial s}{\partial |\gamma|}&\frac{\partial x_0}{\partial \phi}\end{array}\right).
\end{equation}
The minimum MSE of separation is then given by
\begin{equation}
\mathrm{var}(s)\geq(\mathcal{J}_{\boldsymbol{\theta}}\mathcal{Q}_{\boldsymbol{\zeta}}^{-1} \mathcal{J}_{\boldsymbol{\theta}}^{\dagger})_{11}\geq\frac{1}{\mathcal{Q}_{s|\gamma|}}.
\end{equation}

\section{FI for separation using heterodyne detection}
\label{heterodyne}
For heterodyne detection, measurements are performed locally at both telescopes, similar to intensity interferometry, but with a shared phase reference. General error analysis can be used to determine the precision of separation estimation for heterodyne detection. The FI matrix for estimating  $\boldsymbol{\zeta}=(|\gamma|,\phi)^T$ achieved with local heterodyne detection was discussed in Refs.~ \cite{huang2024limited,tsang2011quantum}, giving
\begin{equation}
 \mathcal{F}^{\mathrm{het}}_{\boldsymbol{\zeta}}=\left(\begin{array}{cc}\frac{8\bar{N}^2\gamma^2}{2+6\bar{N}+4\bar{N}^2(1-\gamma^2)}&0\\0&\frac{8\bar{N}^2}{2+6\bar{N}+4\bar{N}^2(1-\gamma^2)}\end{array}\right).
 \label{FI_het}
\end{equation}
According to Eq.~(\ref{error_propagation}), the minimum MSE achieved by heterodyne detection is
\begin{equation}
\mathrm{Var}(s)\geq\frac{1}{(\frac{\partial |\gamma|}{\partial s})^2\mathcal{F}^{\mathrm{het}}_{|\gamma|}+(\frac{\partial \phi}{\partial s})^2\mathcal{F}^{\mathrm{het}}_\phi}=\frac{1}{\mathcal{F}_{{\mathrm{het}}}}.
\label{MSE_het}
\end{equation}

Substituting Eq.~(\ref{FI_het}) into Eq.~(\ref{MSE_het}), we obtain the FI for separation using heterodyne detection
\begin{equation}
\mathcal{F}_{\mathrm{het}}=4\kappa^2\bar{N}^2(q-1)^2q^2\sin(\kappa s)^2+O[\bar{N}]^3\xlongequal{s\to0}0, 
\end{equation}
which is about twice that of intensity interferometry, giving $\mathcal{F}_{\mathrm{het}}/\mathcal{F}_{G^{(2)}}\approx1+\vert \gamma \vert^2\approx2$. Superresolution remains unattainable in heterodyne detection, as the FI vanishes when the separation approaches zero.

\section{The probability distributions of interferometric schemes}
\label{probability}

\begin{figure}[t]
    \centering
    \includegraphics[width=0.47\textwidth]{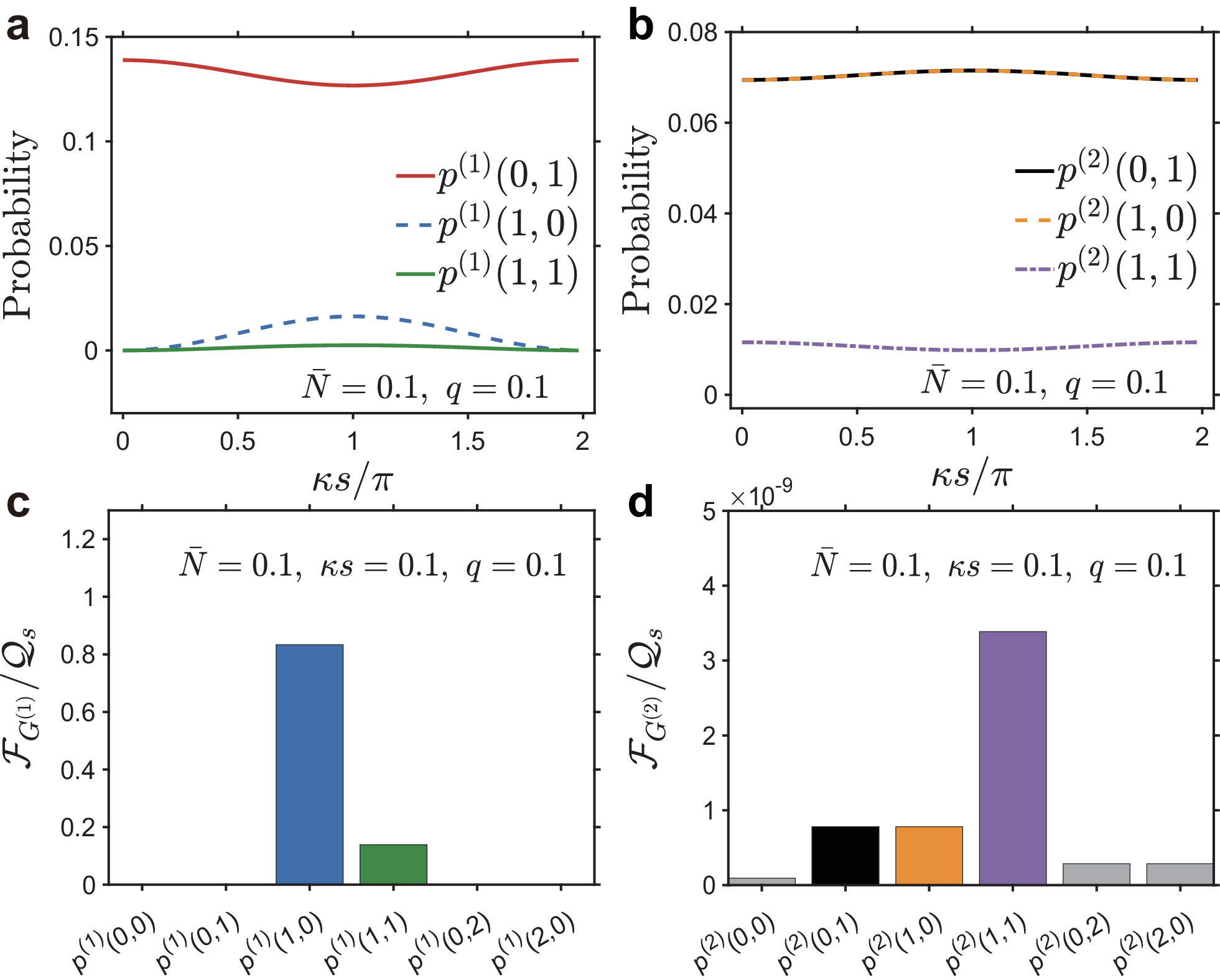}
    \caption{The probability distribution of (a) amplitude interferometry $p^{(1)}(m_1,m_2)$ and (b) intensity interferometry $p^{(2)}(m_1,m_2)$ as a function of the separation $s$ with  $p=0$, $q=0.1$ and $\bar{N}=0.1$. Normalized FI contributed by each measurement outcome for (c) amplitude interferometry and (d) intensity interferometry with $p=0$, $q=0.1$, $\kappa s=0.1$ and $\bar{N}=0.1$.}
    \label{6}
\end{figure}

According to the assumption in Ref.~ \cite{pearce2017optimal}, the states received in the interferometer mode $a_{1,2}$ are given by
\begin{equation}
\begin{aligned}
\label{output state}
&\vert\Psi_{\mathrm{in}}\rangle=\sum_{n_1,n_2}\sqrt{\frac{p_s(n_1,n_2)}{n_1!n_2!}}(\frac{1}{2})^{\frac{n_1+n_2}{2}}\\
&\times\left(\sum_{j=1}^{n_1}\sum_{k=1}^{n_2}C_{n_1}^{j}C_{n_2}^{k}(-1)^ke^{im_1\phi}(\hat{a}_{1}^{\dagger})^{n_1+n_2-j-k}(\hat{a}_{2}^{\dagger})^{j+k}\right)\vert 0 \rangle,
\end{aligned}
\end{equation}
where
\begin{equation}
\begin{aligned}
     p_s(n_1,n_2)&=\frac{\langle n_1 \rangle^{n_1}}{(1+\langle n_1 \rangle)^{n_1+1}}\frac{\langle n_2 \rangle^{n_2}}{(1+\langle n_2 \rangle)^{n_2+1}}\\
\end{aligned}  
\end{equation}
is the photon probability distribution of thermal sources, $\langle n_1 \rangle=\bar{N}(1+|\gamma|), \langle n_2 \rangle=\bar{N}(1-|\gamma|)$ and $\phi=\mathrm{arg}(\gamma)$. For intensity interferometry, the measurement can be described by projecting the telescope state $\vert\Psi_{\mathrm{in}}\rangle$ onto the Fock basis $\vert m_1,m_2\rangle_a$. The probability of getting
outcome $\vert m_1,m_2\rangle_a \langle m_1,m_2\vert_a$ is
\begin{equation}
\begin{aligned}
p^{(2)}(m_1,m_2)=&\sum_{n_{1}}^{m_{1}+m_{2}}\frac{p_s(n_{1},n_{2})}{n_{1}!n_{2}!}\frac{m_1!m_2!}{2^{n_{1}+n_{2}}}\\
&\times\left\vert C_{n_{1}}^{k}\sum_{k}^{m_{2}}C_{n_2}^{m_{2}-k}(-1)^ke^{im_{1}\phi}\right\vert^{2},    
\end{aligned}
\label{p2}
\end{equation}
with $m_1+m_2=n_1+n_2$. In amplitude interferometry, the telescope state undergoes a unitary transformation, which includes the action of a beam splitter (BS) and a phase shift, yielding the final state $\vert \Psi_{\mathrm{out}}\rangle$ in the detection plane for measurement. The action of BS can be described as $\hat{a}_{1}^{\dagger}\to\frac{1}{\sqrt{2}}(\hat{d}_{1}^{\dagger}+\hat{d}_{2}^{\dagger})$ and $\hat{a}_{2}^{\dagger}\to\frac{1}{\sqrt{2}}(\hat{d}_{1}^{\dagger}-\hat{d}_{2}^{\dagger})$. After the BS, a phase shift is imprinted on one of the arms. The total relative phase becomes $\tilde{\phi} = \phi - \alpha$. The measurement can be described by projecting $\vert \Psi_{\mathrm{out}}\rangle$ onto the Fock state basis $\vert m_1,m_2\rangle_d$. The probability of obtaining outcome $\vert m_1,m_2\rangle_d \langle m_1,m_2\vert_d$ is
\begin{equation}
\begin{aligned}
&p^{(1)}(m_1,m_2)=\sum_{n_{1}}^{m_{1}+m_{2}}\frac{p_s(n_{1},n_{2})}{n_{1}!n_{2}!}\frac{m_1!m_2!}{4^{n_{1}+n_{2}}}\\
&\times\left|\sum_{j}^{m_{1}}C_{n_{1}}^{j}C_{n_2}^{m_{1}-j}\left(e^{i\tilde{\phi}}+1\right)^{m_2-n_1+2j}\left(e^{i\tilde{\phi}}-1\right)^{m_1+n
_1-2j}\right|^{2},    
\end{aligned}
\label{p1}
\end{equation} 
with $m_1+m_2=n_1+n_2$. The key differences between amplitude and intensity interferometry are illustrated in Fig.~\ref{6}. For amplitude interferometry, the outcome of destructive interference $p^{(1)}(1,0)$ provides FI that nearly saturates the QFI. In contrast, for intensity interferometry, the two-photon coincidence event $p^{(2)}(1,1)$ contributes the most FI.

\section{The modified probability distribution of interferometric schemes}
\label{relation}
Consider photons from two telescopes now with arbitrary spatial functions $\varphi(\boldsymbol{r})$ and $\zeta(\boldsymbol{r})$, where $\boldsymbol{r}$ is the transverse spatial coordinate. The quantum states for photons collected by two telescopes can be distinguished in the spatial domain, which gives  
\begin{equation}
    |1;\varphi\rangle_{1}=\int d\boldsymbol{r}\varphi(\boldsymbol{r})\hat{a}_1^\dagger(\boldsymbol{r})|0\rangle,~|1;\zeta\rangle_{2}=\int d\boldsymbol{r}\zeta(\boldsymbol{r})\hat{a}_2^\dagger(\boldsymbol{r})|0\rangle.
\end{equation}
The output state of amplitude interferometry is then given by 
\begin{equation}
\label{D2}
\begin{aligned}
  \vert\tilde{\Psi}_{\mathrm{out}}\rangle=&\sum_{n_1,n_2}\sqrt{\frac{p_s(n_1,n_2)}{n_1!n_2!}}(\frac{1}{2})^{n_1+n_2}\\
  &\times\left[(\hat{d}_{1j}^{\dagger}+\hat{d}_{2j}^{\dagger})e^{i\phi}+(\hat{d}_{1k}^{\dagger}-\hat{d}_{2k}^{\dagger})\right]^{n_1}\\
  &\times\left[(\hat{d}_{1j}^{\dagger}+\hat{d}_{2j}^{\dagger})e^{i\phi}-(\hat{d}_{1k}^{\dagger}-\hat{d}_{2k}^{\dagger})\right]^{n_{2}}\vert0\rangle,  
\end{aligned}
\end{equation}
where the subscripts $\{j,k\}$ denote the  photons from different telescopes. The interference visibility can be defined by the spatial overlap of two photons, which is given by
\begin{equation}
    \mathcal{V}=\vert \langle 1;\varphi\vert1;\zeta\rangle \vert= \vert\int \varphi^*(\boldsymbol{r})\zeta(\boldsymbol{r})\hat{a}_1(\boldsymbol{r})\hat{a}_2^\dagger(\boldsymbol{r})d\boldsymbol{r}\vert. 
\end{equation}
To clarify the relation between amplitude interferometry and intensity interferometry, we list the first few conditional probabilities of Eq.~(\ref{D2}), given as
\begin{equation}
    \begin{aligned}
        &\tilde{p}^{(1)}(0,0)=p_s(0,0);\\
        &\tilde{p}^{(1)}(1,0)=\frac12\left[p_s(0,1)(1- \mathcal{V}\cos\tilde{\phi})+p_s(1,0)(1+ \mathcal{V}\cos\tilde{\phi})\right];\\
        &\tilde{p}^{(1)}(0,1)=\frac12\left[p_s(0,1)(1+ \mathcal{V}\cos\tilde{\phi})+p_s(1,0)(1- \mathcal{V}\cos\tilde{\phi})\right];\\
        &\tilde{p}^{(1)}(1,1)=\frac12\{p_s(1,1)[1+\mathcal{V}^2\cos(2\tilde{\phi})]\}\\
        &\quad\quad\quad+\frac14\{[p_s(0,2)+p_s(2,0)][(]2- \mathcal{V}^2-\mathcal{V}^2\cos(2\tilde{\phi})]\}.\\
    \end{aligned}
\end{equation}
The results align with the probability distribution of amplitude interferometry in Eq.~(\ref{p1}) when $\mathcal{V}=1$, and reduce to the probability distribution of intensity interferometry in Eq.~(\ref{p2}) when $\mathcal{V}=0$ or $\tilde{\phi}=\pi/2$. This result demonstrates that intensity interferometry can be viewed as a special case of amplitude interferometry. 

\section{superresolution for interferometry in the single-photon regime}
\label{single-photon regime}
The results for the QFI and amplitude interferometry presented in the main text remain valid in the single-photon regime. In contrast, intensity interferometry is inapplicable since it fundamentally requires coincidence detection between detectors.  Like in Refs.~\cite{wang2021superresolution,tsang2011quantum}, the truncated quantum state can be expressed as 
\begin{equation}
    \begin{aligned}
\rho&=(1-\bar{N})|00\rangle\langle00|+\frac{\bar{N}}{2}\boldsymbol{[}|01\rangle\langle01|+|10\rangle\langle10|\\
&\quad+\gamma^*|01\rangle\langle10|+\gamma|10\rangle\langle01|\boldsymbol{]}+\mathcal{O}\left[\bar{N}\right]^{2} \\
&=(1-\bar{N})\rho_0+\bar{N}\rho_1+\mathcal{O}\left[\bar{N}\right]^{2},
\end{aligned}
\end{equation}
where $\gamma$ is the CDC defined in Eq.~(\ref{gamma}) and the quantum state is expanded in the Fock basis $\vert m_1, m_2 \rangle_a$ with $m_1 + m_2 \leq 1$. The single-qubit mixed state $\rho_1$ encodes the source information. The QFI matrix element for this state can be computed in a basis-independent form as \cite{liu2020quantum}
\begin{equation}
    \mathcal{Q}_{ij}=\mathrm{Tr}\left[(\partial_{i}\rho)(\partial_{j}\rho)\right]+\frac{1}{\mathrm{det}(\rho)}\mathrm{Tr}\left[\rho(\partial_{i}\rho)\rho(\partial_{j}\rho)\right],
\end{equation}
where $\text{Tr}[\cdot]$ and $\text{det}(\cdot)$ denote the trace and determinant, respectively. The relative brightness $q$ is assumed to be predetermined using an independent astronomical technique. Interferometry will then focus on characterizing the spatial properties of the source. When the geometric position $x_0$ is known, the QFI for separation is given by
\begin{equation}
    \begin{aligned}
    &\tilde{\mathcal{Q}}_{s}=\kappa^2\{p^2(4+8(q-1)q)+2[1-(1-2p)^2\cos(\kappa s)]\\
    &\times(q-1)q-8pq^2\}\xlongequal{s\to0}\kappa^2(p^{2}-2pq+q),\\
    \end{aligned}
\end{equation}
which remains constant as the separation approaches zero. This result coincides with Eq.~(\ref{Qs}), apart from a $2\bar{N}$ coefficient. For small separations, the QFI matrix for joint estimation $\boldsymbol{\theta} = (s, x_0)^\top$ becomes
\begin{equation}
   \tilde{\mathcal{Q}}_{\boldsymbol{\theta}}=\kappa^2 \left(\begin{array}{cc}{{p^{2}-2pq+q}}&{{p-q}}\\{{p-q}}&{1}\end{array}\right)+\mathcal{O}\left[s\right]^{2}.
\end{equation}
The fundamental precision, obtained from the inverse of this matrix, also agrees with Eq.~(\ref{Hs}) except for a factor of $2\bar{N}$.

As for the amplitude interferometry, the measurement operator can be expressed as
\begin{equation}
   \Pi_{1}=\frac12 \left(\begin{array}{cc}{1}&{{\mathcal{V}e^{-i\alpha}}}\\{{\mathcal{V}e^{i\alpha}}}&{1}\end{array}\right), \Pi_{2}=\mathbb{I}_2-\Pi_{1},
\end{equation}
where the associated conditional probability probability is $p_j = \mathrm{Tr}[\rho\Pi_j], j={1,2}$. Assuming that interferometry is aligned with $\kappa x_0 = 0$, the FI for separation is given by
\begin{equation}
\begin{aligned}
      \tilde{\mathcal{F}}_{s}&=\frac{\kappa^2 \mathcal{V}^2 [(p-1)q\sin \tilde{\phi}_1-p(q-1)\sin\tilde{\phi}_2]^2}{1-[(]q\mathcal{V}\cos \tilde{\phi}_1-(q-1)\mathcal{V}\cos\tilde{\phi}_2]^2}\\
      &\quad\xlongequal[\mathcal{V}\to1,\alpha\to0]{s\to0}\kappa^2(p^{2}-2pq+q),
\end{aligned}
\end{equation}
where $\tilde{\phi}_1=(p-1)\kappa s+\alpha$ and $\tilde{\phi}_2= p \kappa s+\alpha$. This constant precision demonstrates the superresolution capability of amplitude interferometry in the single-photon regime, which extends the results of Ref.~\cite{lupo2020quantum} to sources with unequal brightnesses. Furthermore, the nulling strategy remains quantum-optimal performance for small separations in the single-photon regime.

%\bibliography{apssamp}

\end{document}